\magnification = \magstep0
\hsize=18.61truecm  \hoffset=-1.0truecm  \vsize=24.4truecm  \voffset=-0.3truecm

\font\sc=cmr9
\font\ssc=phvr   scaled 700 
\font\sscap=phvb scaled 700 
\font\cap=phvb scaled 900

\font\titl=phvb  scaled 2000
\font\stitl=phvb  scaled 1980
\font\au=phvr    scaled 1200
\font\sic=phvro  scaled 700

\def\sglbaselines{\baselineskip=10pt    \lineskip=0pt   \lineskiplimit=0pt}
\def\smlbaselines{\baselineskip=9pt     \lineskip=0pt   \lineskiplimit=0pt}

\def\vsl{\vskip\baselineskip}   \def\vs{\vskip 5pt} 
\parindent=5pt \nopagenumbers

\def\omit#1{\empty}
\def\ba{\kern -4pt}  \def\baa{\kern -6pt}
\parskip = 0pt 
\def\ts{\thinspace} \def\cl{\centerline}
\def\ni{\noindent}  
\def\nhi{\noindent \hangindent=0.9truecm}

\def\makeheadline{\vbox to 0pt{\vskip-30pt\line{\vbox to8.5pt{}\the
                               \headline}\vss}\nointerlineskip}

\def\footnoterule{\kern-5pt \hrule width \fullhsize \kern 2.6pt \vskip 1pt}
\output={\plainoutput}    \pretolerance=10000   \tolerance=10000

\def\gapprox{$_>\atop{^\sim}$} 
          
\newdimen\sa  \def\sd{\sa=.1em \ifmmode $\rlap{.}$''$\kern -\sa$
                               \else \rlap{.}$''$\kern -\sa\fi}
\newdimen\sb  \def\md{\sb=.02em\ifmmode $\rlap{.}$'$\kern -\sb$
                               \else \rlap{.}$'$\kern -\sb\fi}

\def\ss{\ifmmode ^{\prime\prime}$\kern-\sa$ \else $^{\prime\prime}$\kern-\sa\fi}
\def\mm{\ifmmode ^{\prime}$\kern-\sa$ \else $^{\prime}$\kern-\sa \fi}

\sc

\headline{{\bf Nature Letter, 20 January 2011, 000, 000--000 \hfill\null}}

\cl{\null}\vskip -\baselineskip \vskip -9pt

\noindent{\titl Supermassive black holes do not correlate with galaxy disks or pseudobulges}

\vsl\vskip -3pt

\sglbaselines

\ni {\au John Kormendy\footnote{$^1$}{\ssc\baa Department of Astronomy, University of Texas at Austin,
                                               1 University Station, Austin, TX 78712-0259, USA},
R.~Bender\footnote{$^2$}{\ssc\baa Max-Planck-Institut f\"ur Extraterrestrische Physik, 
                                               Giessenbachstrasse, D-85748 Garching-bei-M\"unchen, Germany}$^,$
\ba\footnote{$^3$}{\ssc\baa Universit\"ats-Sternwarte, Scheinerstrasse 1, D-81679 M\"unchen, Germany},
\& M.~E.~Cornell$^1$
}
\vsl\vskip -8pt

\vskip 17.36truecm

\smlbaselines

\ni {\sscap
Figure 1 | Correlations of dynamically measured black hole masses M$_\bullet$ with the K-band (2.2 $\mu$m) 
absolute magnitude of (a) the disk component with bulge light removed, (b) the bulge with disk light removed, and 
(c) the pseudobulge with disk light removed.   All plotted data are published elsewhere;
parameters and sources are discussed in the Supplementary Information, and those for disk galaxies
are tabulated~there.  Elliptical galaxies are plotted in black; classical bulges are plotted in red; 
pseudobulges are plotted in blue.  One galaxy with a dominant pseudobulge but with a possible small classical 
bulge (NGC 2787) is plotted with a blue symbol that has a red center.  In least-squares fits, it is 
included with the pseudobulges.  Error bars are 1 sigma.    In panel (b) the red and black points show a
good correlation between M$_\bullet$ and bulge luminosity; a symmetric, least-squares fit\raise2pt\hbox{4} 
of a straight line has $\chi$\raise2pt\hbox{2} = 12.1 per degree of freedom and a Pearson correlation coefficient 
of r = -0.82.  (All $\chi$\raise2pt\hbox{2} values quoted in this paper are per degree of freedom.)  In contrast,
in  panel (a), the red and blue points together confirm a previous result\raise2pt\hbox{1} that BHs do not correlate 
with disks: $\chi$\raise2pt\hbox{2} = 81 and r = 0.41.  Green points are for galaxies that contain
neither a classical nor a pseudo bulge but only a nuclear star cluster; i.{\ts}e., these are pure-disk galaxies.
They are not included in the above fit, but they strengthen our conclusion.  Similarly, in panel (c) the blue 
points for pseudobulges show no correlation: $\chi$\raise2pt\hbox{2} = 63 and r = 0.27.  In all panels, galaxies that 
have only M$_\bullet$ limits are plotted with open symbols; they were chosen to increase our dynamic range.  They, 
too, support our conclusions.  This figure uses K-band magnitudes to minimize effects of star formation and internal
absorption, but the Supplementary Information shows that Figure 1 looks essentially the same for V-band (5500 A) magnitudes.
}
\vskip -21.0truecm


\newdimen\fullhsize
\fullhsize=18.6 cm 
\hsize=9.05 cm
\def\fullline{\hbox to \fullhsize}
\let\lr=L \newbox\leftcolumn
\output={\if L\lr
         \global\setbox\leftcolumn=\columnbox \global \let\lr=R
         \else  \doubleformat \global \let\lr=L\fi
         \ifnum\outputpenalty>-2000 \else \dosupereject\fi}
\def\doubleformat{\shipout\vbox{\makeheadline
         \fullline{\box\leftcolumn\hfil\columnbox}
         \makefootline}
\advancepageno}
\def\columnbox{\leftline{\pagebody}}


\sglbaselines

\omit{
{\cap The masses of supermassive black holes are known~to correlate with the properties of the bulge 
components of their host galaxies\raise2pt\hbox{1-5}.  In contrast, they appear not to correlate with galaxy 
disks\raise2pt\hbox{1}.  Disk-grown pseudobulges are intermediate in properties between bulges and disks\raise2pt\hbox{6}; 
it is unclear whether they do\raise2pt\hbox{1,5} or do not\raise2pt\hbox{7-9} correlate with black holes
in the same way that bulges do.  At stake are conclusions about which parts of galaxies coevolve with 
black holes\raise2pt\hbox{10}, possibly by being regulated by energy feedback from black holes\raise2pt\hbox{11}.
Here we confirm that black holes do not correlate with disks and show that they correlate little or not at 
all with pseudobulges.  Our results are based in part on recent measurements of velocity dispersions in the biggest 
bulgeless galaxies\raise2pt\hbox{12} and on new bulge-pseudobulge classifications.  We suggest that there are two 
different modes of black hole feeding.  Black holes in bulges grow rapidly to high masses when mergers drive 
gas infall that feeds quasar-like events.  In contrast, small black holes in bulgeless galaxies and galaxies
with pseudobulges grow as low-level Seyferts.  Growth of the former is driven by global processes, so the 
biggest black holes coevolve with bulges, but growth of the latter is driven locally and stochastically, 
and they do not coevolve with disks and pseudobulges.
}
}

{\cap The masses of supermassive black holes are known~to correlate with the properties of the bulge 
components of their host galaxies\raise2pt\hbox{1-5}.  In contrast, they appear not to correlate with galaxy 
disks\raise2pt\hbox{1}.  Disk-grown pseudobulges are intermediate in properties between bulges and disks\raise2pt\hbox{6}.
It has been unclear whether they do\raise2pt\hbox{1,5} or do not\raise2pt\hbox{7-9} correlate with black holes in 
the same way that bulges do, because too few pseudobulges were classified to provide a clear result.  At stake are 
conclusions about which parts of galaxies coevolve with black holes\raise2pt\hbox{10}, 
possibly by being regulated by energy feedback from black holes\raise2pt\hbox{11}.
Here we report pseudobulge classifications for
galaxies with dynamically detected black holes and combine them with recent
measurements of velocity dispersions in the biggest bulgeless galaxies\raise2pt\hbox{12}.
These data confirm that black holes do not correlate with disks and show that they correlate 
little or not at all with pseudobulges. 
We suggest that there are two 
different modes of black hole feeding.  Black holes in bulges grow rapidly to high masses when mergers drive 
gas infall that feeds quasar-like events.  In contrast, small black holes in bulgeless galaxies and galaxies
with pseudobulges grow as low-level Seyferts.  Growth of the former is driven by global processes, so the 
biggest black holes coevolve with bulges, but growth of the latter is driven locally and stochastically, 
and they do not coevolve with disks and pseudobulges.
}

\vskip 3pt

\hfuzz=20pt

      The well known correlation$^{1,5}$ between dynamically measured BH masses M$_\bullet$ and the absolute magnitudes 
of elliptical galaxies (black points) and the bulge parts of disk galaxies (red points) is shown in Figure 1(b).  
The correlation has $\chi^2$ = 12 per degree of freedom implying moderate intrinsic scatter.  This result and a tighter 
correlation$^{2-5}$ between M$_\bullet$ and host velocity dispersion~$\sigma$ (Figure 2) motivate the idea that BHs and 
host bulges evolve together and regulate each other$^{10,11}$.  All new results in this paper are in contrast to these
two correlations.

      Figure 1(a) plots $M_\bullet$ versus the absolute magnitude of only the disk part of the host galaxy, with the
bulge luminosity removed.  We conclude that BHs do not correlate with galaxy disks.  This confirms an earlier conclusion$^{1}$ 
based on the more indirect observation that BHs do not correlate with total (bulge{\ts}+{\ts}disk) luminosities of 
disk galaxies. In Figure 1(a), the color~of~the points{\ts}--{\ts}which encodes bulge type (see below){\ts}--{\ts}is 
irrelevant.  A least-squares fit to the red and blue points has correlation coefficient $r$ = 0.41. One-sigma 
errors imply $\chi^2$~=~81 per degree of freedom: the data do not respect~the~weak~anticorrelation.  The green points
for pure-disk (that is, completely bulgeless) galaxies further confirm the large scatter and lack of correlation.

      Figure 1(c) plots $M_\bullet$ versus $M_K$ for ``pseudobulge'' components with disk light removed.  
They are also included in (b) in ghostly 

\vfill

\vskip 0.7truein

\cl{\null}

\vskip 0.7truein

\vfill

\cl{\null}

\vskip 22.5pt

\noindent light blue to show how they compare with classical bulges
and ellipticals.  Pseudobulges required explanation, as follows.

      Much work over several decades has shown that the high-stellar-density central components in
disk galaxies -- all of which we used to call ``bulges'' -- come in two varieties.  How to distinguish them
is discussed in the Supplementary Information.   The difference was first found observationally but is now 
understood to be a result of fundamentally different formation mechanisms$^{6}$.

\vfill

\includegraphics{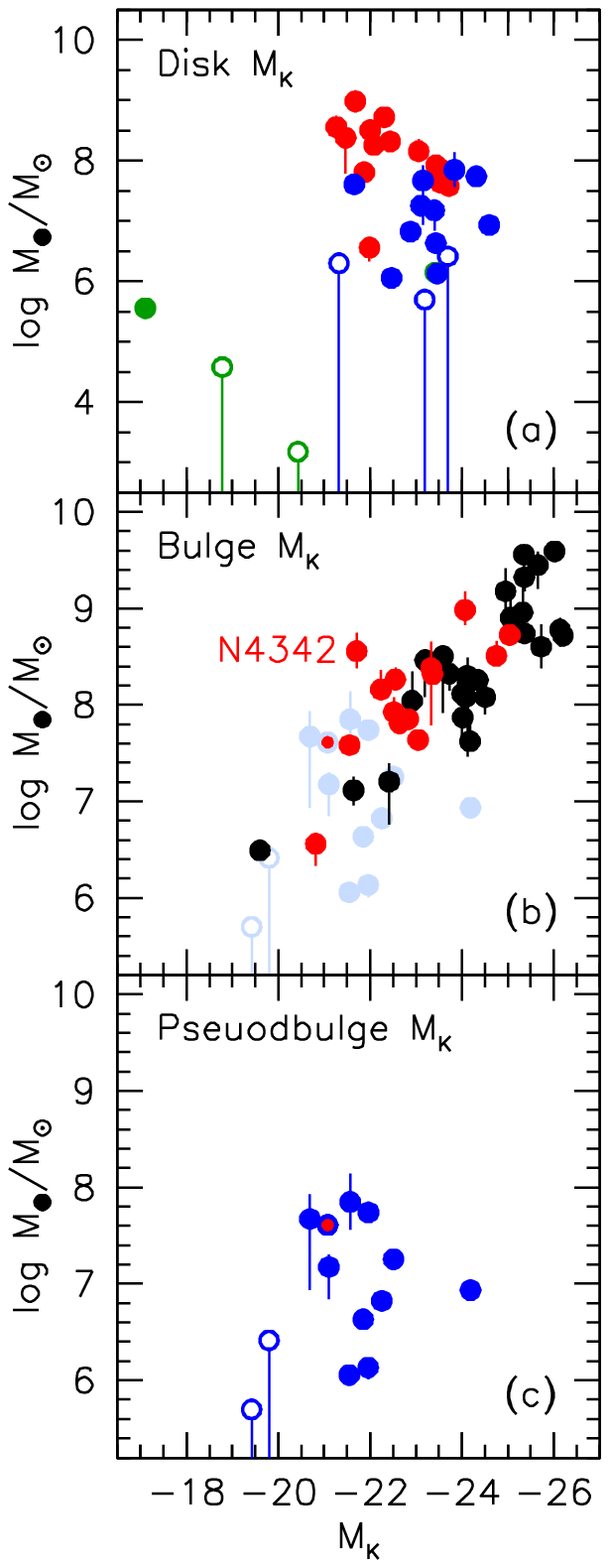}

\cl{\null}

\eject

      Classical bulges (red points in the Figures) are indistinguishable from ellipticals in their structure,
velocity distributions, and parameters.  Our well developed paradigm is that they formed by galaxy
mergers$^{13}$ in our hierarchically clustering universe.   Mergers are discrete events separated by long 
``dead times'', while they happen, their time scales are short; i.{\ts}e.,\ts$\approx$ the crossing time.  
Gravitational torques scramble disks into ellipticals$^{13}$ and dump large quantities of 
gas into the center.  Observations$^{14}$ and theory$^{15}$ suggest that the process feeds both starbursts 
and BHs and causes the latter to grow rapidly in quasar-like events.  

      Pseudobulges$^{6}$ (blue points in the Figures) are observed to be more disk-like than
classical bulges.  They are believed to form more gently by the gradual internal redistribution
of angular momentum in quiescent~galaxy~disks.  The driving agents are nonaxisymmetries such 
as bars.  One result is the gradual buildup of a high-density central component that can
be recognized (Supplementary Information) because it remains rather disk-like.  We call
these components ``pseudobulges'' to emphasize their different formation mechanism without
forgetting that they look superficially like -- and are commonly confused with -- classical bulges. 
The difference from bulges that is most relevant here is this: Gradual gas infall 
may provide less BH feeding and may drive slower BH growth.  One purpose of this paper is to 
contrast BH\ts--{\ts}bulge and BH\ts-- {\ts}pseudobulge correlations to look for clues about BH 
growth mechanisms and the consequent coevolution (or not) of BHs with host galaxies.

      With this context, we can interpret Figure 1.    Galaxies that contain classical bulges  
are consistent with the correlations for elliptical galaxies except for one discrepant object (the 
bulge-dominated S0 galaxy NGC 4342).~The implication is that classical bulges and ellipticals 
coevolve with BHs in the same way.  For that coevolution, it is irrelevant that bulges are now
surrounded by disks whereas ellipticals are not.

      We reach a different conclusion for pseudobulges based on new classifications and
\hbox{measurements of pseudobulge-to-total} luminosity ratios for all disk galaxies in our sample$^5$ that~have
dynamical BH detections (Kormendy, in preparation; see the Supplementary Information for a list).  A conservative
interpretation of Figure 1(b) is that they are roughly consistent with the correlation for classical bulges and 
ellipticals but have much more scatter.  In particular, some pseudobulges deviate from the correlation for 
ellipticals in having smaller BHs.  This was not seen in some previous work$^{1,5}$ because samples were small 
and because many pseudobulges in BH galaxies had not been classified.  But, as published samples have grown 
larger, the hints have grown stronger that pseudobulges do not correlate with BHs in the same way as classical
bulges$^{7-9}$.  We confirm these hints.  Particularly compelling is the fact that our galaxies and a new 
sample of BH detections based on water masers$^{9}$ have no overlap and independently lead to the same conclusion.

       Figure 1(c) shows the pseudobulges without guidance from the red and black points.  The sample is small,
but we have enough dynamic range to conclude that we see no correlation~at~all.  The cumulative amount of BH
growth is not extremely different in classical and pseudo bulges, but there is no sign in the correlations that 
BH feeding has affected the pseudobulges.

      The second and more compelling BH{\ts}--{\ts}host galaxy correlation~is the one between M$_\bullet$ and the 
velocity dispersion $\sigma$ of the stars at radii where they do not feel the BH gravitationally$^{2-5}$.
Here~$\sigma$ is averaged inside the ``effective radius'' r\lower2pt\hbox{e} that contains half of the bulge light.  
Figure 2 shows this correlation.

      As is well known, ellipticals and classical bulges share the same tight correlation.  But as in Figure 1, 
pseudobulges at best show a much larger scatter (Figure 2a).  Without the guidance of the red and black points 
(Figure 2b), they show essentially no correlation.  Larger samples that reach smaller M$_\bullet$ may 
show a weak relationship$^{16-20}$.  But we conclude that classical bulges and pseudobulges show very different 
correlations with M$_\bullet$.  The ones for classical bulges are tight enough to suggest coevolution.  
Whether pseudobulges correlate with M$_\bullet$ with large scatter or not at all, the weakness of any 
correlation (r = $-$0.08 here) makes no  compelling case that pseudobulges and BHs coevolve, beyond the obvious
expectation that it is easier to grow bigger BHs and bigger pseudobulges in bigger galaxies that contain more fuel.

\headline{{\bf Nature, 20 January 2011, 000, 000--000 \hfill\null}\rm 2}

\cl{\null}

\vskip 4.2truein

\includegraphics{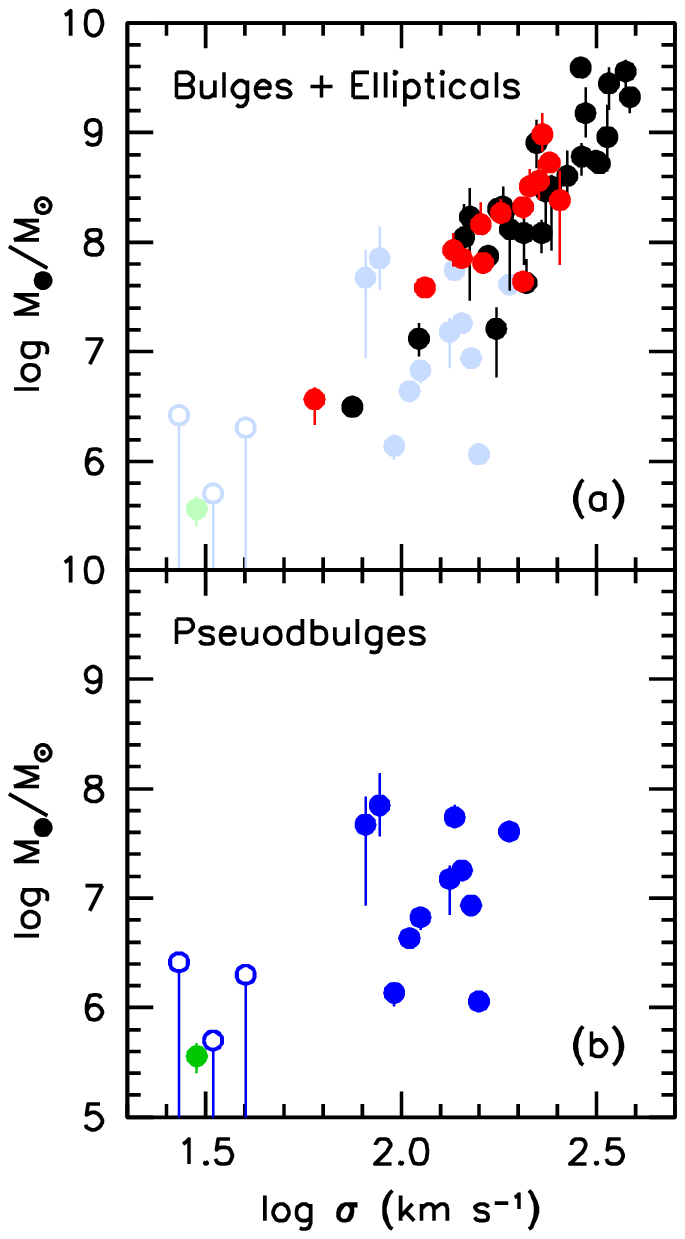}

\ni {\sscap Figure 2 | Correlation of dynamically measured BH masses M$_\bullet$ with the velocity dispersion 
$\sigma$ of the host elliptical (black points), classical bulge (red points), pseudobulge (blue points), or
nuclear star cluster (green point).   Date sources are given in the Supplementary Information.  Error bars are 
1 sigma.  The red and black points show the well known M$_\bullet$ -- $\sigma$ correlation\raise2pt\hbox{2--5}; 
$\chi$\raise2pt\hbox{2} = 5.0 per degree of freedom and r = 0.89.  Reducing $\chi$\raise2pt\hbox{2} to 1.0 implies 
that the intrinsic scatter in log M$_\bullet$ at fixed $\sigma$ is 0.26, consistent with previous 
derivations\raise2pt\hbox{4,5}.  This is the tightest correlation between BHs and host galaxy properties and the 
one that most motivates the idea that BHs and bulges coevolve.  In contrast, the blue points for pseudobulges 
show no correlation: $\chi$\raise2pt\hbox{2} = 10.4; r = -0.08.~This extends suggestions\raise2pt\hbox{7--9}
that the BH -- $\sigma$ relation is different for pseudobulges than it is for classical bulges and elliptical galaxies.
}

\vs

      From the point of view of galaxy formation by hierarchical clustering,
pseudobulge galaxies are already pure-disk galaxies$^{12}$.  What about even more extreme
galaxies that contain neither a classical nor a pseudo bulge?  At least some of them contain~BHs.
Some have active galactic nuclei (AGNs), although these are rare in bulgeless galaxies$^{21}$.~And 
BHs with M$_\bullet$ = 10$^4$\ts--\ts10$^6${\ts}M$_\odot$ have confidently been discovered in 
(pseudo)bulge-less galaxies$^{18,21}$.  The most extreme example is NGC 4395, an Sm galaxy that
contains only a tiny, globular-cluster-like nucleus but that has an AGN powered by a BH$^{22}$
with M$_\bullet$ = (3.6 $\pm$ 1.1) $\times$ 10$^5$~M$_\odot$ well measured by reverberation mapping$^{23}$.
It is the green point with the small error bar in Figure 2.
Other such objects include the Sd galaxy NGC 3621$^{24}$ and the Sph galaxy POX 52,
which contains an AGN powered by a BH of mass M$_\bullet \simeq$ 10$^5$ M$_\odot^{18,25}$.
Also, it is likely that low-M$_\bullet$ AGN samples include bulgeless galaxies,$^{16-19}$ although some 
are distant and not well resolved. 

      The lack of correlation of BHs with disks and pseudobulges plus the discovery of BHs in 
(pseudo)bulgeless galaxies together are critical clues to black hole feeding mechanisms.  They motivate 
the following hypothesis.

\lineskip=-15pt \lineskiplimit=-15pt

      We suggest that there are two fundamentally different feeding  mechanisms for BHs:

      1 -- 
      The traditional mechanism is rapid feeding during major mergers when large amounts of gas
fall into galaxy centers.  
We follow other authors$^{14,15}$ who suggest that, at some, perhaps~late stage in the merger, the BH grows quickly 
in a \hbox{quasar-like event.}  These are the growth episodes that dominate the waste mass budget of quasar activity$^{15,26}$ 
that is well explained by the observed density of the biggest BHs in the universe$^{27}$.  In this mode, BH and galaxy 
growth are controlled by the same global processes and these, we suggest, result in BH -- bulge coevolution.  
If BHs and galaxy formation ever regulate each other$^{11}$, this is the likely scenario in which it happens.

      2 -- 
      In contrast, the nuclear activity in bulgeless galaxies is, by and large, weak$^{21}$.  Big pseudobulges 
(NGC 1068 is the highest-luminosity one in Figure 1) can host classical Seyfert nuclei, but there is
no sign that these affect global galaxy structure.  NGC 1068 is a prototypical oval galaxy$^{6}$ with 
a prominent pseudobulge that, we believe, formed slowly by inward gas transport in spite of episodic nuclear activity.  
We suggest that the second mode of BH growth is such weak nuclear activity driven stochastically by local processes 
that feed gas from $\sim$\ts10\raise2pt\hbox{2} pc where it makes pseudobulges in to the BH.  The processes are not 
understood in detail.  But exactly this kind of feeding mode is proposed and modeled in (26); the models 
suggest that this kind of feeding does not affect galaxy formation.  Differences in BH growth and coevolution with
host galaxies have also been proposed in studies of the lowest-M$_\bullet$ BHs detected mainly
via AGN activity$^{16,18}$.  The same scenario can also be reached by studying AGN demographics$^{28}$. 
Other recent papers$^{29,30}$ further explore local processes of BH feeding.
Figures 1~and~2 tell us that this mode involves little or no coevolution of BHs with any component of the host galaxy.  

      The seed BHs that grew into today's supermassive BHs are~not securely identified.  But the smallest BHs grown 
via local feeding plausibly  remain most like those seeds.  Since mergers of disk galaxies are believed to make bulges 
and ellipticals, we suggest that small BHs that are grown by local processes are the seeds of the generally 
larger BHs that are grown in part by mergers.

\headline{{\bf Nature, 20 January 2011, 000, 000--000 \hfill\null}\rm 3}

\vs\vskip 1pt

{\ni\sscap Received 12 July 2010.} 

\vs\vskip 2pt

\smlbaselines

{\frenchspacing

\ssc

\nhi 1.\quad Kormendy, J., \& Gebhardt, K.~Supermassive black holes in galactic nuclei.
             in {\sic 20$^{\rm th}$ Texas Symposium on Relativistic Astrophysics}. (eds Wheeler, J. C., \& 
             Martel, H.) 363--381 (AIP, 2001).

\nhi 2.\quad Ferrarese, L., \& Merritt, D.~A fundamental relation between supermassive black holes and their
              host balaxies.~{\sic Astrophys.~J.}~{\sscap 539}, L9--L12 (2000).

\nhi 3.\quad Gebhardt, K., {\sic et al.}~A relationship between nuclear black hole mass and galaxy velocity 
             dispersion.~{\sic Astrophys.~J.}~{\sscap 539}, L13--L16 (2000).

\nhi 4.\quad Tremaine, S. {\sic et al.} The slope of the black hole mass versus velocity dispersion correlation.
             {\sic Astrophys.~J.}~{\sscap 574}, 740--753 (2002).

\nhi 5.\quad G\"ultekin, K., {\sic et al.}~The M--$\sigma$ and M--L relations in galactic bulges,
               and determinations of their intrinsic scatter. {\sic Astrophys.~J.}~{\sscap 698},
               198--221 (2009).

\nhi 6.\quad Kormendy, J., \& Kennicutt, R.~C.~Secular evolution and the formation of pseudobulges
              in disk galaxies.~{\sic Annu.~Rev.~Astron.~Astrophys.}~{\sscap 42}, 603--683 (2004).

\nhi 7.\quad  Hu, J.~The black hole mass -- stellar velocity dispersion correlation: bulges versus
               pseudo-bulges.~{\sic Mon.~Not.~R.~Astron.~Soc.}~{\sscap 386}, 2242--2252 (2008).

\nhi 8.\quad Nowak, N., Thomas, J., Erwin, P., Saglia, R. P., Bender, R., \& Davies, R. I.
             Do black hole masses scale with classical bulge luminosities only?  The case of the two 
             composite pseudo-bulge galaxies NGC 3368 and NGC 3489.
             {\sic Mon.~Not.~R.~Astron.~Soc.}~{\sscap 403}, 646--672 (2010).

\nhi 9.\quad Greene, J. E., {\sic et al.} Precise black hole masses from megamaser disks: Black hole -- bulge
               relations at low mass. {\sic Astrophys.~J.}~{\sscap 721}, 26--45 (2010).

\nhi 10.\quad Ho, L.~C., Ed.~{\sic Carnegie Observatories Astrophysics Series,
             Volume 1: Coevolution of Black Holes and Galaxies} (Cambridge Univ.~Press, 2004).

\nhi 11.\quad Silk, J., \& Rees, M.~J. Quasars and galaxy formation. {\sic Astron. Astrophys.} 
              {\sscap 331}, L1--L4 (1998).

\nhi 12.\quad Kormendy, J., Drory, N., Bender, R., \& Cornell, M.~E.~Bulgeless giant galaxies
               challenge our picture of galaxy formation by hierarchical clustering.~{\sic
               Astrophys.~J.} {\sscap 723}, 54--80 (2010).

\nhi 13.\quad Toomre, A.~Mergers and some consequences.~in {\sic  Evolution of Galaxies and Stellar 
               Populations} (eds Tinsley, B. M. \& Larson, R. B.) 401--426 (Yale University Observatory, 1977).

\nhi 14.\quad Sanders, D.~B., Soifer, B.~T., Elias, J.~H., Madore, B.~F., Matthews, K.,
               Neugebauer, G., \& Scoville, N.~Z.~Ultraluminous infrared galaxies and the
               origin of quasars.~{\sic Astrophys.~J.}~{\sscap 325}, 74--91 (1988).

\nhi 15.\quad Hopkins, P.~F., Hernquist, L., Cox, T.~J., di Matteo, T., Robertson, B.,
               \& Springel, V.~A unified, merger-driven model of the origin of starbursts, 
               quasars, the cosmic X-ray background, supermassive black holes, and galaxy 
               spheroids.~{\sic Astrophys.~J.~Suppl.~Ser.}~{\sscap 163}, 1--49 (2006).

\nhi 16.\quad Barth, A. J., Greene, J. E., \& Ho, L. C. Dwarf Seyfert 1 nuclei and the low-mass end of the 
               M\lower2pt\hbox{BH} -- $\sigma$ relation. {\sic Astrophys.~J.} {\sscap 619}, L151--L154 (2005).

\nhi 17.\quad Greene, J. E., \& Ho, L. C. The M\lower2pt\hbox{BH}\ts--\ts$\sigma_*$ relation in local active galaxies.
                {\sic Astrophys.~J.} {\sscap 641}, L21--L24 (2006).

\nhi 18.\quad Greene, J. E., Ho, L. C., \& Barth, A. J. Black holes in pseudobulges and spheroidals: A change in the
               black hole -- bulge scaling relations at low mass. {\sic Astrophys.~J.} {\sscap 688}, 159--179 (2008).

\nhi 19.\quad Bentz, M. C., Peterson, B. M., Pogge, R. W., \& Vestergaard, M. The black hole mass -- bulge luminosity
              relationship for active galactic nuclei from reverberation mapping and {\sic Hubble Space Telescope\/} imaging.
              {\sic Astrophys.~J.} {\sscap 694}, L166--L170 (2009).

\nhi 20.\quad Woo, J.-H., {\sic et al.} The Lick AGN monitoring project: The M\lower2pt\hbox{BH} -- $\sigma_*$ relation
              for reverberation-mapped active galaxies. {\sic Astrophys.~J.} {\sscap 716}, 269--280 (2010).

\nhi 21.\quad Ho, L. C. Nuclear activity in nearby galaxies. {\sic Annu.~Rev.~Astron.~Astrophys.}~{\sscap 46}, 475--539 (2008).

\nhi 22. \quad Filippenko, A. V., \& Ho, L. C.	A low-mass central black hole in the bulgeless Seyfert 1 galaxy NGC 4395.
                {\sic Astrophys.~J.} {\sscap 588}, L13--L16 (2003).

\nhi 23.\quad Peterson, B. M., {\sic et al.} Multiwavelength monitoring of the dwarf Seyfert 1 galaxy NGC 4395. I. 
               A Reverberation-based measurement of the black hole mass. {\sic Astrophys.~J.} {\sscap 632}, 799--808 (2005).

\nhi 24.\quad Barth, A. J., Strigari, L. E., Bentz, M. C., Greene, J. E., \& Ho, L. C. Dynamical constraints on the masses of the 
               nuclear star cluster and black hole in the late-type spiral galaxy NGC 3621.
               {\sic Astrophys.~J.} {\sscap 690}, 1031--1044 (2009).

\nhi 25.\quad Thornton, C. E., Barth, A. J.; Ho, L. C.; Rutledge, R. E., \& Greene, J. E.
              The host galaxy and central engine of the dwarf active galactic nucleus POX 52.
              {\sic Astrophys.~J.} {\sscap 686}, 892--910 (2008).

\nhi 26.\quad Hopkins, P. F., \& Hernquist, L. Fueling low-level AGN activity through stochastic accretion of cold gas.
                 {\sic Astrophys.~J.~Suppl.~Ser.}~{\sscap 166}, 1--36 (2006).

\nhi 27.\quad Yu, Q. \& Tremaine, S. Observational constraints on growth of massive black holes.
              {\sic Mon.~Not.~R.~Astron.~Soc.}~{\sscap 335}, 965--976 (2002).

\nhi 28.\quad Schawinski, K., {\sic et al.} Galaxy zoo: The fundamentally different co-evolution of supermassive
                 black holes and their early- and late-type host galaxies. {\sic Astrophys.~J.}~{\sscap 711}, 284--302 (2010).

\nhi 29.\quad Kumar, P., \& Johnson, J. L. Supernovae-induced accretion and star formation in the inner kilparsec
              of a gaseous disc.
              {\sic Mon.~Not.~R.~Astron.~Soc.}~{\sscap 404}, 2170--2176 (2010).

\nhi 30.\quad Hopkins, P. F., \& Quataert, E. How do massive black holes get their gas?
              {\sic Mon.~Not.~R.~Astron.~Soc.} {\sscap 407}, 1529--1564 (2010).

\vskip 1.0truein

}

\vsl

\vsl

\vskip -1.2truein

{\ni\sscap Supplementary Information} {\ssc is linked to the online version of the paper 
                                        at www.nature.com/nature.}

\vskip 4pt

{\ni\sscap Acknowledgments} {\ssc We acknowledge with pleasure the collaboration of Niv
Drory on work$^{12}$ leading up to this paper.  We thank Niv and Jenny Greene for 
helpful comments on the MS and Jenny for communicating the
maser BH detection results before publication$^{9}$.  We also thank Karl Gebhardt for permission 
to use M$_\bullet$ for NGC 4736 and NGC 4826 and John Jardel for permission to use his updated 
M$_\bullet$ for NGC 4594 before publication.  Some data used here were 
obtained with the Hobby-Eberly Telescope (HET).  It is a joint project of the University of Texas at Austin, 
Pennsylvania~State~University, Stanford University Ludwig-Maximilians-Universit\"at Munich, and 
Georg-August-Universit\"at G\"ottingen.  The HET is named in honor of its principal benefactors,
William P.~Hobby and Robert E.~Eberly.  We made extensive use of data from the Two Micron All 
Sky Survey, a joint project of the University of Massachusetts and the Infrared Processing 
and Analysis Center/California Institute of Technology funded by NASA and by the NSF.   We also made extensive 
use of the NASA/IPAC Extragalactic Database (NED), which is operated by Caltech and JPL under contract 
with NASA, of the HyperLeda database $<$http://leda.univ-lyon1.fr$>$, and of NASA's Astrophysics 
Data System bibliographic services.  Finally, we are grateful to the National Science Foundation for
grant support.
\lineskip=-15pt \lineskiplimit=-15pt
}

\vskip 4pt

{\ni\sscap Author Contributions} {\ssc\frenchspacing J.K. led the program, carried out the analyis for this
paper and wrote most of the text.  M.E.C. oversaw the HET observations, preprocessed the HET spectra, and 
provided technical support throughout the project.  R.B. calculated the velocity dispersions from the HET spectra
and made all least-squares fits.  All authors contributed to the writing of this paper.}

\vskip 4pt

{\ni\sscap Author Information} {\ssc Reprints and permissions information is available at
www.nature.com/reprints.  The authors declare no competing financial interests.
Correspondence and requests for materials should be addressed to J.K. (kormendy@astro.as.utexas.edu).



\eject

\cl{\null}
\vskip 0pt

\noindent{\stitl Supplementary Information}

\headline{{\bf Nature, 20 January 2011, 000, 000--000 \hfill\null}\rm 4}

\vsl

\rm \sglbaselines

      The first section summarizes the differences between classical bulges and pseudobulges
and lists the criteria that we use to distinguish them.  The second section tabulates the data
used in the construction of the figures in the main text.  The third section shows a V-band
version of Figure 1.

\vs
\cl{\bf Internal Secular Evolution of Disk Galaxies and the}
\cl{\bf Distinction Between Classical and Pseudo Bulges}
\vs

\headline{{\bf Nature, 20 January 2011, 000, 000--000 \hfill\null}\rm 4}

\lineskip=-20pt \lineskiplimit=-20pt
  
      The distinction between classical bulges and pseudobulges plays a critical role in the  main text.
It is not yet securely a part of the extragalactic folklore, so we review it briefly:
\vskip 4pt

      {\bf Classical bulges:} Our standard picture for the formation of elliptical galaxies
by major galaxy mergers in the context of hierarchical clustering was summarized in the main text. 
Classical bulges have observed properties that are similar to those of comparably low-luminosity
ellipticals and are believed to have formed like ellipticals in major mergers.  Renzini$^{31}$ clearly 
states the usual interpretation of classical bulges: ``It appears legitimate to look at bulges as ellipticals 
that happen to have a prominent disk around them [and] ellipticals as bulges that for some reason 
have missed the opportunity to acquire or maintain a prominent disk.''  This story is well known.
\vskip 4pt

      {\bf Pseudobulges} are less well known.  Kormendy \& Kennicutt review$^{6}$ a large
body of evidence that, between major merger events, galaxy disks evolve continually
and slowly (``secularly'') as nonaxisymmetries transfer angular momentum outward.  The main driving agents
are bars, globally oval disks$^{6,32,33}$, and global spiral structure that reaches the 
center of the galaxy.  The result is a variety of well known structural features such as outer rings,
inner rings that encircle the ends of bars, and pseudobulges.  The latter were first recognized$^{33,34}$
because they are more disk-like than are classical bulges.  More recent work confirms and greatly extends this
distinction$^{6}$.  In parallel, theoretical work has shown that disks fundamentally rearrange their angular 
momentum distributions in a way that dumps large quantities of gas into the center.  There, it is expected and observed 
to feed starbursts and to build high-density components -- we infer: the observed pseudobulges -- that were made slowly 
out of disk material and not quickly in a merger.  The distinction is important here because mergers are thought 
to feed BHs rapidly$^{14,15}$.  It therefore becomes important to ask whether these two very different ways to 
form bulge-like central components do or do not grow BHs similarly.  We confirm previous suggestions$^{7-9}$
that they do not.  In fact, we find no BH -- pseudobulge correlations at all.  This suggests that BHs coevolve only 
with classical~bulges and ellipticals\ts--{\ts}i.{\ts}e., only with remnants of major mergers.

      The key to recognizing pseudobulges is that they ``remember'' their disky origins.  They
continue to have structural, dynamical, and star formation properties that are more like those of their
associated disks than they are like those of ellipticals and classical bulges.  The criteria used to classify
bulges as pseudo or classical are listed in the Kormendy \& Kennicutt review$^{6}$ and in the paper
in preparation on BH hosts.  Here is a brief summary; justification is given in the above papers: 

\vskip 5truein

\cl{\null}

\vskip 0.2truein
\vsl

      1 -- Pseudobulges often have disky structure -- their apparent flattening 
is similar to that of the outer disk, or they contain spiral structure that reaches 
all the way in to the galaxy center.  Classical bulges look much rounder than their disks
unless the galaxy is seen face-on.  They cannot show spiral structure.

      2 -- In relatively face-on galaxies the presence of a nuclear bar implies a
pseudobulge.  Bars are purely disk phenomena.  

      3 -- In edge-on galaxies, boxy bulges are edge-on bars; seeing one is sufficient 
grounds for identifying a pseudobulge. 

      4 -- Most pseudobulges have S\'ersic$^{35}$ index $n < 2$, whereas almost all classical 
bulges have $n \geq 2$. Here $n$ is a parameter of the projected brightness profile,
log I $\propto$ radius\raise2pt\hbox{1/n}.

      5 -- In pseudobulges, ordered motions (rotation) are slightly more important with
respect to random motions ($\sigma$) than they are in classical bulges.  This is even
more true in disks.

      6 -- Many pseudobulges are low-$\sigma$ outliers in the well known correlation$^{36}$
between (pseudo)bulge luminosity and $\sigma$.  Frequently, $\sigma$ decreases inward in the 
central few arcsec.

      7 -- If the center of the galaxy is dominated by gas, dust, and star formation
but there is no sign of a merger in progress, then the bulge is at least mostly pseudo.  

      8 -- Small bulge-to-total luminosity ratios do not guarantee that a bulge
is pseudo, but if B/T\ts\gapprox 0.5, then the bulge is classical.

\vs
\cl{\bf Table of Data for Disk Galaxies Shown in Figures}
\vs

      Supplementary Table 1 lists the data for the disk galaxies that are included in Figures 1 and 2.
It is an excerpt from a more detailed table in a paper (Kormendy, in preparation) on BH host 
galaxies.  Most references for parameter sources are given there; exceptions are as follows:
The galaxy sample, BH masses, and parameters not otherwise credited are from the recent study 
of M$_\bullet$ correlations by G\"ultekin$^{5}$.  They include $\sigma$, which is tabulated in 
G\"ultekin's paper as the mean velocity dispersion inside approximately the ``effective radius''
r\lower2pt\hbox{e} that contains half of the light of the bulge.  Note: Only disk galaxies are 
included in the table.  Parameters for ellipticals are given in G\"ultekin's paper.
BH mass limits in essentially (pseudo)bulge-less galaxies are from paper (12).

      The table is divided into two parts, galaxies with classical bulges and then galaxies
with pseudobulges.  Column (8) lists which of the criteria in the previous section
led to the bulge classifications in Columns (6) and (7).

      Photometric decomposition of brightness distributions~into (pseudo)bulge and disk
components is discussed~for~each galaxy in the paper in preparation on BH hosts.
Fifteen of the 25 galaxies in the table have~$(P)B/T$~measurements from 2 -- 7 independent
sources.  Ten have only one~source; seven of these are measured in the BH hosts paper.
A variety of techniques are used in published papers; most of them use S\'ersic functions
for bulges and exponentials for disks.  The decompositions in the BH hosts paper are based
on composite brightness profiles from as many sources as possible.  Most of them are measured
in the above paper.  They include HST profiles whenever possible.  No $(P)B/T$ ratio is 
seriously compromised by the lack of HST data.  The BH hosts paper uses one-dimensional
S\'ersic-exponential decompositions of major-axis profiles but takes ellipticity profiles
fully into account in calculating $(P)B/T$.  

\vskip 5truein

\cl{\null}

\vskip 10truein

\includegraphics{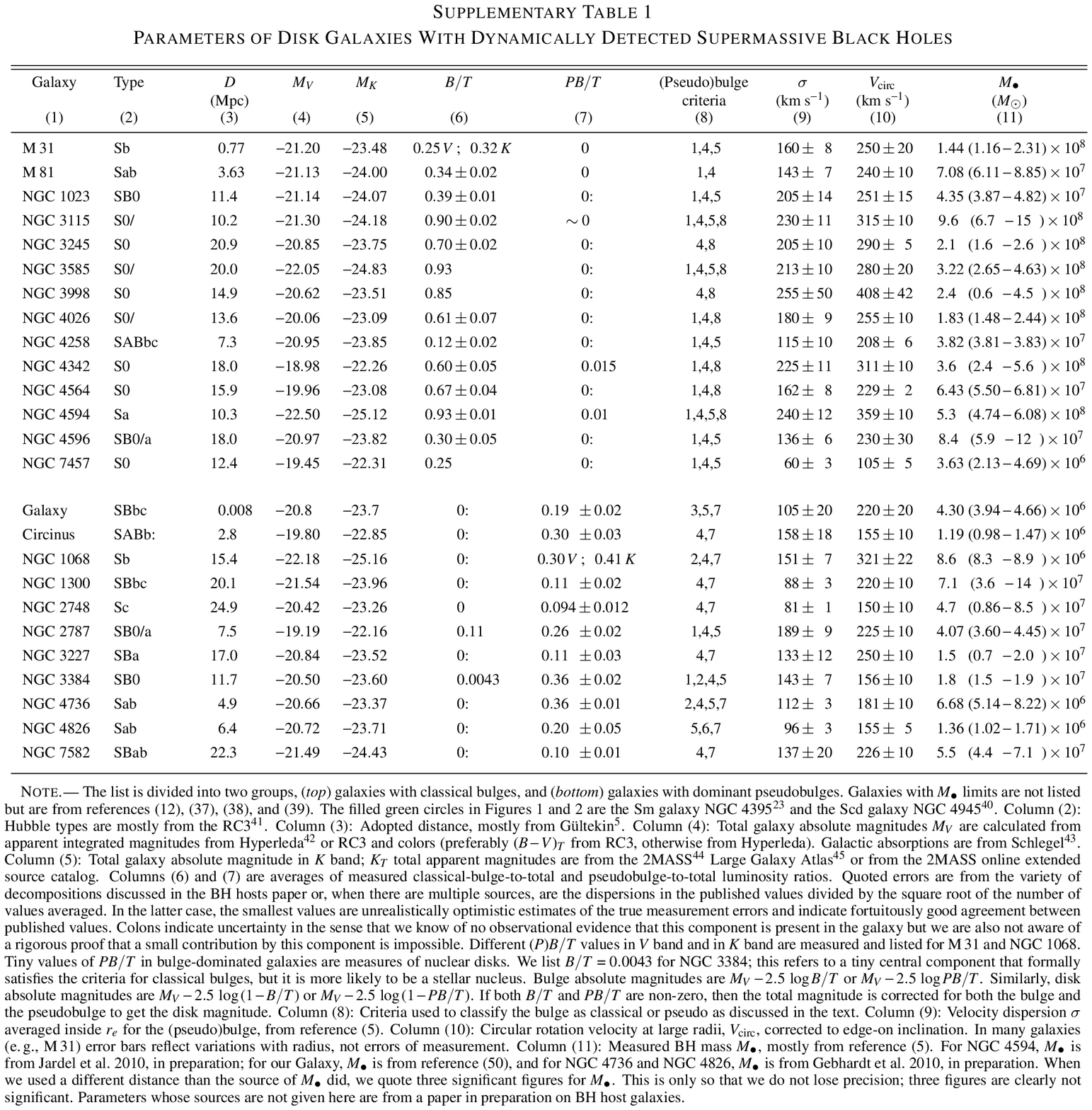}

\cl{\null}

\headline{{\bf Nature, 20 January 2011, 000, 000--000 \hfill\null}\rm 5}

\eject

\vfill

\vfill

      For giant ellipticals, published estimates of M$_\bullet$ that are based on stellar dynamics 
need correction for the effects of triaxiality and for the effect of including halo dark matter (DM) 
in the models.  These corrections are mostly relevant only for the biggest ellipticals, i.{\ts}e.,
the ones that we believe were formed by dry mergers$^{46}$.  Including DM inevitably 
increases M$_\bullet$ estimates, and so far, including triaxiality has had the same effect.  
Corrected M$_\bullet$ values are available for three ellipticals:
NGC 3379$^{47}$,
NGC 4486$^{48}$, and
NGC 4649$^{49}$.
The corrected masses are factors of $\sim$\ts2 larger than those used in the figures.  This does not 
effect our conclusions.  But it artificially increases the scatter.  Until more corrections are available,
we decided to be consistent and used the G\"ultekin tabulated masses$^{5}$.

\vs
\cl{\bf V-Band Version of Figure 1}
\vs

      Supplementary Figure 1 shows a version of Figure 1 that is based on optical (V-band) absolute
magnitudes rather than infrared absolute magnitudes.  We noted in the Figure 1 caption that 
we used K-band magnitudes there because they minimize effects of star formation and internal absorption.  
However, the present figure based on V-band magnitudes looks virtually identical to Figure 1 and leads to the 
same conclusions.  This means that variations in mass-to-light ratios for these galaxies, some or 
which are actively forming stars and others of which are not, do not affect our results.

\vfill



\includegraphics{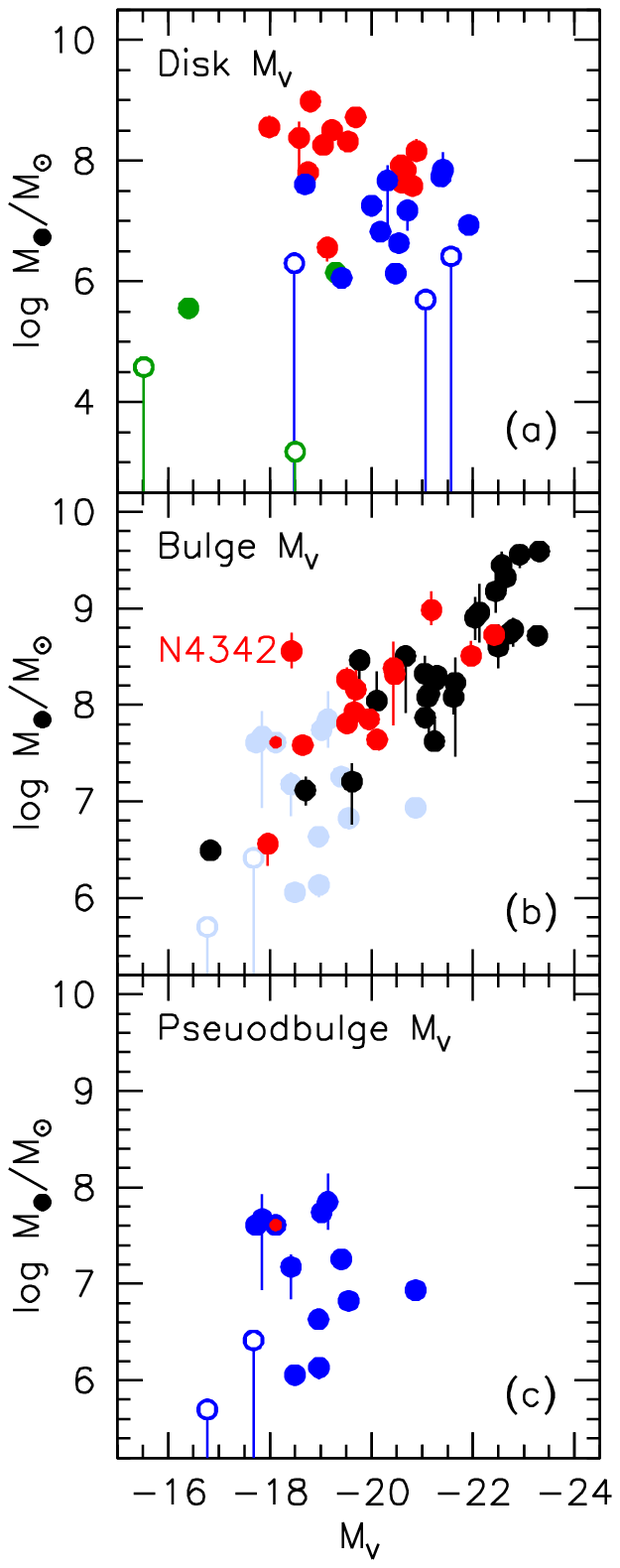}

\ni {\sscap Supplementary Figure 1 | Correlations of dynamically measured BH mass M$_\bullet$ with 
(a) disk, (b) classical bulge, and (c) pseudobulge absolute magnitude in V band (5500 A).   Elliptical 
galaxies are plotted with black points; classical bulges are plotted in red, and pseudobulges are plotted 
in blue.  Panel (b) also shows the pseudobulges in light blue.  Green points are for galaxies that 
contain neither a classical nor a pseudo bulge but only a nuclear star cluster.  Open circles show galaxies 
with BH mass limits.
}

\eject

\vs\vs
\cl{\bf Acknowledgments}
\vs

      This work was supported by the National Science Foundation under grant AST-0607490.

\vs\vs
\cl{\bf References}
\vs

\smlbaselines

{\frenchspacing

\ssc

\nhi 31.\quad  Renzini, A. 1999. Origin of bulges. in {\sic The Formation of Galactic Bulges}, (eds. Carollo, C. M., 
              Ferguson, H. C., \& Wyse, R. F. G.) 9--25 (Cambridge University Press, 1999)

\nhi 32.\quad Kormendy, J., \& Norman, C. A. Observational constraints on driving mechanisms for spiral density waves.
              {\sic Astrophys.~J.}~{\sscap 233}, 539--552 (1979).

\nhi 33.\quad Kormendy, J. Observations of galaxy structure and dynamics. in {\sic 12\raise2pt\hbox{th} Saas-Fee Course,
         Morphology and dynamics of galaxies.}, (eds. Martinet, L., \& Mayor, M.) 113--288 (Geneva Observatory, 1982)

\nhi 34.\quad Kormendy, J. Kinematics of extragalactic bulges: Evidence that some bulges are really disks.
               in {\sic IAU Symposium 153, Galactic Bulges} (eds. Dejonghe, H., \& Habing, H. J.) 209--228 (Kluwer, 1993)

\nhi 35.\quad S\'ersic, J.~L. {\sic Atlas de Galaxias Australes}. (Observatorio Astron\'omico, Universidad Nacional de C\'ordoba, 1968)

\nhi 36.\quad Faber, S. M., \& Jackson, R. E. Velocity dispersions and mass-to-light ratios for elliptical galaxies. 
              {\sic Astrophys.~J.}~{\sscap 204}, 668--683 (1976).

\nhi 37.\quad Gebhardt, K., {\sic et al.}~M{\ts}33: A galaxy with no supermassive black hole. {\sic
              Astron.~J.}~{\sscap 122}, 2469--2476  (2001).

\nhi 38.\quad B\"oker, T., van der Marel, R.~P., \& Vacca, W.~D.~CO band head spectroscopy
        of IC 342: Mass and age of the nuclear star cluster.~{\sic Astron.~J.}~{\sscap 118}, 831--842 (1999).

\nhi 39.\quad Valluri, M., Ferrarese, L., Merritt, D., \& Joseph, C. L. The low end of the supermassive black hole mass function: 
              Constraining the mass of a nuclear black hole in NGC 205 via stellar kinematics.
              {\sic Astrophys.~J.}~{\sscap 628}, 137--152 (2005).

\nhi 40.\quad Greenhill, L., J., Moran, J. M., \& Herrnstein, J. R. The distribution of H$_2$O maser emission in the nucleus of NGC 4945.
              {\sic Astrophys.~J.}~{\sscap 481}, L23--L26 (1997).

\nhi 41.\quad de Vaucouleurs, G., de Vaucouleurs, A., Corwin, H. G., Buta, R. J., Paturel, G., \& Fouqu\'e, P.
               {\sic Third Reference Catalogue of Bright Galaxies}. (Springer, 1991).

\nhi 42.\quad Paturel, G., Petit, C., Prugniel, Ph., Theureau, G., Rousseau, J., Brouty, M.,
             Dubois, P., \& Cambr\'esy, L.~HYPERLEDA. I. Identification and designation of galaxies.
             {\sic Astron.~Astrophys.}~{\sscap 412}, 45--55 (2003).

\nhi 43.\quad Schlegel, D. J., Finkbeiner, D. P., \& Davis, M. Maps of dust infrared emission for use in estimation of 
               reddening and cosmic microwave background radiation foregrounds. {\sic Astrophys.~J.}~{\sscap 500}, 525--553 (1998).

\nhi 44.\quad Skrutskie, M.~F., {\sic et al.} The two micron all sky survey (2MASS).
              {\sic Astron.~J.}~{\sscap 131}, 1163--1183 (2006).

\nhi 45.\quad Jarrett, T. H., Chester, T., Cutri, R., Schneider, S. E., \& Huchra, J. P. The 2MASS Large Galaxy Atlas.
               {\sic Astron.~J.}~{\sscap 125}, 525--554 (2003).

\nhi 46.\quad Kormendy, J., Fisher, D. B., Cornell, M. E., \& Bender, R. Structure and formation of elliptical 
                and spheroidal galaxies. {\sic Astrophys.~J.~Suppl.~Ser.}~{\sscap 182}, 216--309 (2009).

\nhi 47.\quad van den Bosch, R. C. E., \& de Zeeuw, P. T. Estimating black hole masses in triaxial galaxies.
              {\sic Mon. Not. R. Astron. Soc.} {\sscap 401}, 1770--1780 (2010). 

\nhi 48.\quad Gebhardt, K., \& Thomas, J. The black hole mass, stellar mass-to-light ratio, and dark
              halo in M{\ts}87. {\sic Astrophys.~J.}~{\sscap 700}, 1690--1701 (2009).

\nhi 49.\quad Shen, J., \& Gebhardt, K. The supermassive black hole and dark matter halo of NGC 4649 (M{\ts}60).
              {\sic Astrophys.~J.}~{\sscap 711}, 484--494 (2010).

\nhi 50.\quad Genzel, R., Eisenhauer, F., \& Gillessen, S. The massive black hole and nuclear star cluster in
              the center of the Milky Way. {\sic Rev.~Mod.~Phys.}, in press;
              preprint at $<$http://arxiv.org/abs/1006.0064$>$ (2010).  

\headline{{\bf Nature, 20 January 2011, 000, 000--000 \hfill\null}\rm 6}

\end